\documentclass[aps,prl,twocolumn,showpacs,preprintnumbers,amsmath,amssymb]{revtex4}
\usepackage[english]{babel}

\usepackage{natbib}
\usepackage{graphicx}
\usepackage{color}
\usepackage[colorlinks,bookmarks=false,citecolor=blue,linkcolor=red,urlcolor=blue]{hyperref}
\usepackage{times}

\def\opone{\le\textbf{}\textbf{}avevmode\hbox{\small1\kern-3.8pt\normalsize1}}
\def\hk{hyperkagome }

\begin{document}
\author{E.~J.~Bergholtz}
\affiliation{Max-Planck-Institut f\"{u}r Physik komplexer Systeme, N\"{o}thnitzer Stra{\ss}e 38, D-01187 Dresden, Germany}
\author{A.~M.~L\"auchli}
\affiliation{Max-Planck-Institut f\"{u}r Physik komplexer Systeme, N\"{o}thnitzer Stra{\ss}e 38, D-01187 Dresden, Germany}
\author{R. Moessner}
\affiliation{Max-Planck-Institut f\"{u}r Physik komplexer Systeme, N\"{o}thnitzer Stra{\ss}e 38, D-01187 Dresden, Germany}

\title{Symmetry Breaking on the Three-Dimensional Hyperkagome Lattice of Na$_4$Ir$_3$O$_8$ }

\date{\today}

\begin{abstract} We study the antiferromagnetic spin-$1/2$ Heisenberg model on the highly frustrated, three-dimensional, hyperkagome lattice of Na$_4$Ir$_3$O$_8$ using a series expansion method. We propose a valence bond crystal with a $72$ site unit cell as a ground state that supports many, very low lying, singlet excitations. Low energy spinons and triplons are confined to emergent lower-dimensional motifs. Here, and for analogous kagome and pyrochlore states, we suggest finite temperature signatures, including an Ising transition, in the magnetic specific heat due to a multistep breaking of discrete symmetries.
\end{abstract}


\maketitle

\paragraph{Introduction.---} In the last decades, the search for exotic magnetic states has mainly concentrated on low-dimensional systems, $d \leq 2$, where the effects of fluctuations are particularly strong \cite{anderson,ms,rk,ml}. It has only been realised more recently that---whereas magnetic order is more easily destabilised in low dimension---certain unusual phases are in fact more robust in $d = 3$ \cite{u1}.

\begin{figure}[t]
\centerline{\includegraphics[width=1.0\linewidth]{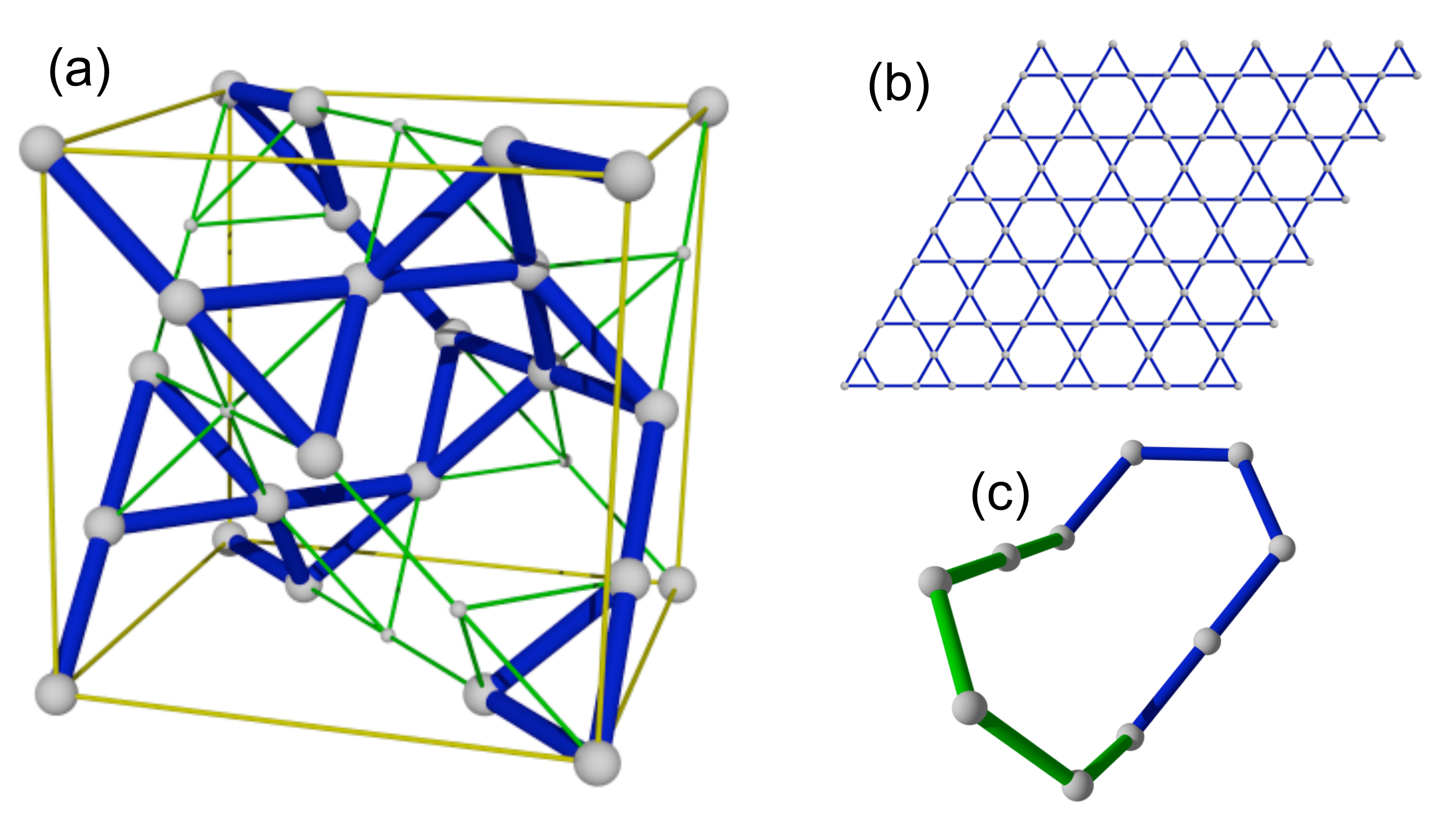}}
\caption{(Color online) Lattice structure:  (a) cubic unit cell of the \hk lattice (thin bonds and small sites show the extra bonds and sites of the pyrochlore lattice).
(b) the 2D kagome lattice. (c) a 'decagon', the smallest loop of the \hk lattice, highlighting the structure of two intersecting planes by different coloring.
\label{fig:uc}}
\end{figure}

A case currently under active investigation is the compound Na$_4$Ir$_3$O$_8$ \cite{exp}, where a subtle interplay of spin and orbital degrees of freedom leads to an effective $S=1/2$ description \cite{gang} of the magnetic Ir ions occupying the sites of a \hk lattice of corner-sharing triangles (Fig. \ref{fig:uc}). This material shows no sign of magnetic ordering down to the lowest temperatures, thus presenting a candidate for a quantum spin liquid \cite{lawler1,zhou,lawler2}. Here, we follow a different lead, namely the possibility of the magnetic state being a valence bond crystal (VBC), breaking translational but not spin-rotational invariance. Interestingly the gadolinium gallium garnet lattice~\cite{GGG} with nearest neighbor interactions consists of two independent \hk networks of opposite chirality, providing a broader scope for our VBC proposal.

We find a candidate ground state with a huge (in this case 72-site) unit cell. We elucidate its structure and discuss its excitations, which include emergent low-dimensional motifs. We discuss the general implications of these large-unit-cell VBCs for the expected (sequence of) phase transitions as the temperature is lowered. 
These considerations also apply to analogous states proposed for kagome~\cite{mz,sh,sh2} and pyrochlore~\cite{berg,msg,hbb,tsunet} magnets.
We also address some general features of the series expansion method \cite{sh,sh2} used to derive our effective Hamiltonian.

\paragraph{Lattice structure and effective model.---}
The \hk lattice [Fig.~\ref{fig:uc}(a)] is a three-dimensional lattice of corner sharing triangles and as such a cousin of the kagome lattice in $d = 2$ shown in Fig.~\ref{fig:uc}(b). It can be obtained from the pyrochlore lattice by removing one site per tetrahedron, yielding a connected network of triangles with a unit cell of twelve sites in which all sites are equivalent~\cite{exp}. We set the length of the cubic unit cell to one in the following. An important property is that the shortest cycle (closed loop) on this lattice beyond the triangles involves ten bonds [cf.~Fig.~\ref{fig:uc}(c)] and will be called "decagon" in the following,  in analogy with the length six "hexagon" loops on kagome. 
   
We aim to determine the quantum mechanical ground state of the antiferromagnetic nearest-neighbor $S=1/2$ Heisenberg model
on the \hk lattice with $N$ sites and periodic boundary conditions given by the Hamiltonian:
\begin{equation}
\label{eqn:ham}
H= J \sum_{\langle i,j \rangle}\mathbf{S}_i\cdot \mathbf{S}_j\ .
\end{equation} 
Focusing on valence bond crystal formation, we perform series expansions around arbitrary nearest-neighbor singlet (dimer) coverings of the \hk lattice 
to obtain an effective Hamiltonian (energy function) for the $2^{N/3+1}$ dimer coverings, following earlier work of Singh and Huse on the kagome lattice~\cite{sh,sh2}.
We subsequently determine and characterize the dimer coverings minimizing the energy. 
While the triangles covered by a dimer
are local ground states of Hamiltonian~\eqref{eqn:ham}, the $N/6$ triangles not hosting a dimer induce quantum fluctuations and if strong enough can melt the dimerised state.

We start with an arbitrary dimer covering, $\mathcal D$, of 'strong' singlet bonds and treat the remaining part of $H$ as a perturbation: 
$$H\rightarrow H_\lambda= J \sum_{\langle i,j \rangle\in \mathcal D}\mathbf{S}_i\cdot \mathbf{S}_j+\lambda J \sum_{\langle i,j \rangle\not\in \mathcal D}\mathbf{S}_i\cdot \mathbf{S}_j.$$ This procedure folds the quantum fluctuations of the
empty triangles into effective potential energy terms (given by power series in $\lambda$) for various local motifs appearing in the dimer coverings. If a VBC is actually favored---or is at least locally stable---such  an expansion should be well behaved, converging towards the correct ground state and energy thereof at reasonably low expansion orders also at $\lambda=1$.

In our analysis, we expand the Hamiltonian up to fifth order in $\lambda$. To this order, the Hamiltonian consists of a constant term, a term favoring two empty triangles linked by one dimer and a term favoring five empty triangles arranged along decagons. The effective model is illustrated in the following:
\begin{equation}
\label{eqn:eff_ham}
\includegraphics*[width=0.95\linewidth]{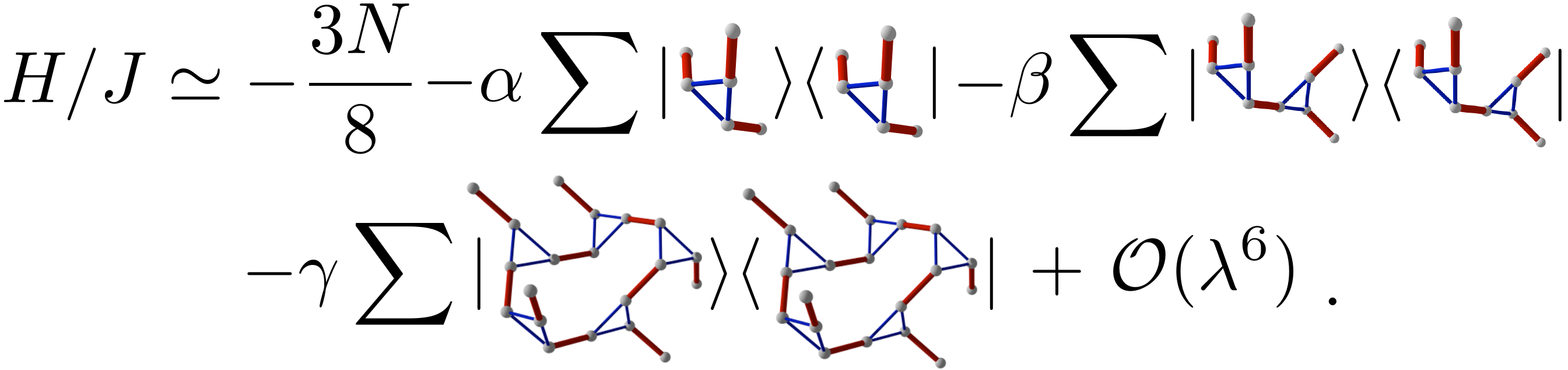}
\end{equation}
The coefficients are given as  
$\alpha=9\lambda^2/32+ 15  \lambda^4/2048+ \lambda^5/128$, 
$\beta= 5 \lambda^4/264 + \lambda^5/2048$
and
$\gamma=105 \lambda^5/8192$. Note that since the number of empty triangles is the same for all dimer coverings, the term proportional to $\alpha$ simply corresponds to a global energy shift.

\paragraph{Minimization of the effective Hamiltonian.---} Given the constrained nature of dimer coverings, minimizing the classical dimer Hamiltonian (\ref{eqn:eff_ham}) among all $N_{\mathcal D}=2^{N/3 + 1}$ dimer coverings of the \hk lattice is a non-trivial task. In order to find ground states we resort to a classical dimer Monte Carlo approach based on non-local worm updates~\cite{SandvikMoessnerDimers}. Simulating the effective
classical dimer model~(\ref{eqn:eff_ham}) at $\lambda=1$ on a range of cubic samples of $N=L^3\times12$ sites with $L=3,4,\ldots,9,12,15,18$ we found states which reach the lower bound for the energy
of the term binding two neighboring empty triangles (an empty triangle can have at most three neighboring empty triangles). 
On an $L=6$ sample we uncovered our lowest energy state state (to be described below) which satisfies all two-triangle interactions and has $144$ satisfied decagon terms.
Simulations involving only decagon energy terms reached a slightly larger number of 147 perfect decagons on the same sample. So it seems---unlike the case of the kagome lattice---that there is no way of minimizing the shortest closed loop term and the neighboring empty triangle interaction simultaneously on the \hk lattice. Given that the energy penalty is quite small (3 out of 147) and the decagon coupling $\gamma$ is smaller than the empty triangle binding term $\beta$, we are optimistic that the reported VBC state is the bona fide ground state of the classical dimer model Eq.~(\ref{eqn:eff_ham}).

\begin{table}
\begin{tabular}{|c|c|c|c|c||c|}
    \hline
 Order & 0, 1 & 2, 3  & 4 & 5 & SL    
  \\ \hline
  Energy  & $-0.375$  & $-0.421875$ & $-0.427979$ & $-0.430115$ & $-0.424$ \\
    \hline

\end{tabular}
\caption{VBC energies per site in units of $J$ for the infinite system evaluated at $\lambda=1$ for the \hk versus the lowest spin liquid (SL) energy quoted in the literature \cite{lawler2}. 
}
 \label{energies}
\end{table}

As shown in Table \ref{energies} the candidate VBC on the \hk does very well in minimizing the energy as compared to the best spin liquid estimate given in the literature \cite{lawler2}. In fact, it seems to compare even more favorably to the spin liquid than is the case in  two dimensions (comparing the results obtained in Refs.~\onlinecite{sh2,ran}). On general grounds one may also expect that a VBC state is likely to be more stable on a lattice with longer closed loops. To higher order in $\lambda$, increasing numbers of terms involving geometrical motifs growing in size will enter the effective Hamiltonian, which will generically
compete with each other. 
We note that sub-leading terms tend to increase the magnitude of the coefficient of pairwise empty triangle interactions, $\beta$, while the coefficient of the closed loop terms, $\gamma$, decreases at sub-leading order. Thus, it is likely that the regime where our candidate state wins is maintained at higher orders. 

We have also established that the series expansion itself has the capability of signaling its own breakdown by explicitly carrying out the analysis on the checkerboard lattice where long range quadrumer order (rather than dimer order) is expected \cite{Fouet}. In this case the series favor different ground state symmetries already at the two first non-trivial orders (second and third in this case) when evaluated at $\lambda=1$. This is in sharp contrast to the kagome and \hk expansions. Moreover, it was demonstrated explicitly that the energy of the candidate VBC changes only little even between fifth \cite{sh} and eighth \cite{sh2} order on the kagome lattice.

\paragraph{Description of the VBC state.---}
\begin{figure*}[h!t]
\centerline{\includegraphics[width=1.0\linewidth]{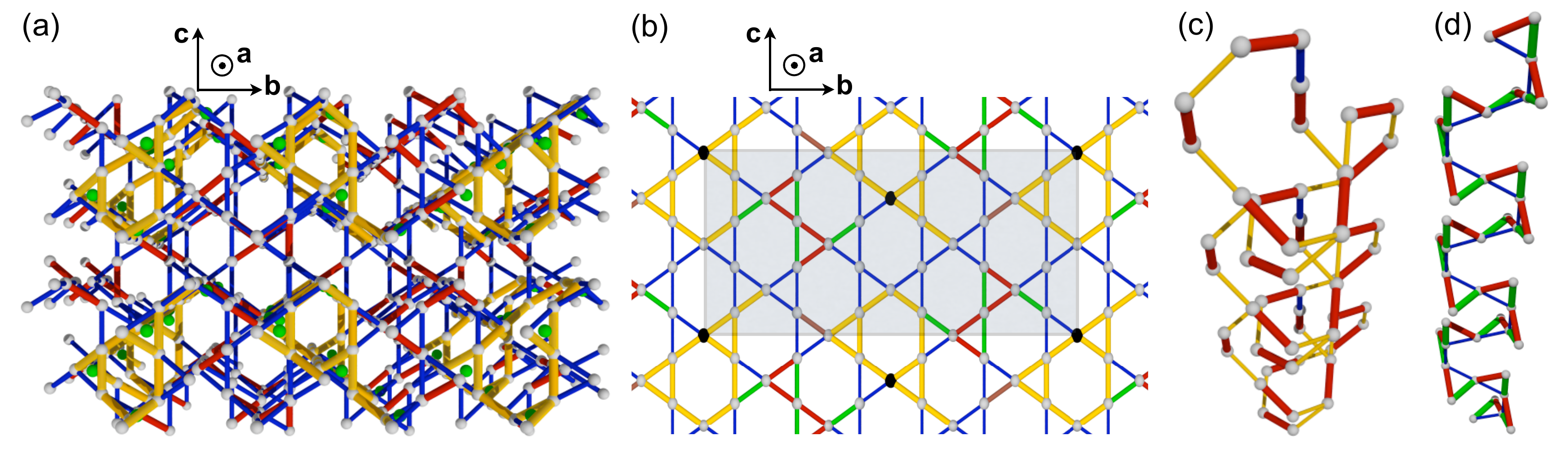}}
\caption{(Color online) VBC structure: (a) details of the VBC ground state on a slab of the \hk viewed along a $[110]$ direction: dimers are red and weak bonds blue except on the decagon-rods, which are entirely yellow. The empty triangles are indicated by green spheres. 
The two-dimensional, kagome-like, projection of the state is shown in (b), highlighting the positions and orientations of the decagon-rods (yellow) and the $\Delta$-chains (red {\it vs} green). The shaded region corresponds to one choice of a VBC unit cell. 
The positions of the backbones of the decagon-rods are marked by black and their bridging dimers are indicated in brown.  
(c) shows a decagon rod---the main building block of the VBC ground state. The decagon-rods are directed along a backbone (blue).
The two degenerate dimer coverings of a $\Delta$-chain (red {\it vs} green) are shown in (d). 
 \label{fig:gs}}
\end{figure*}

Our proposed VBC state has a unit cell of 72 sites~\footnote{An orthorombic unit cell of the VBC is given by the vectors $\mathbf{a}=(1,1,0),\ \mathbf{b}=(0,0,3),\ \mathbf{c}=(1,-1,0)$.}
 (containing 36 dimers) and is illustrated in Fig.~\ref{fig:gs}(a). 
The two main building blocks are:
i) two kinds of {\em decagon-rods} running along a $[110]$ direction each consisting of a chain of stacked perfect decagons, with two neighboring decagons sharing a dimer and two empty triangles [cf.~Fig.~\ref{fig:gs}(c)] 
and ii) $\Delta$-chains ---with two degenerate dimer coverings for each chain---running parallel to the rods, see Fig.~\ref{fig:gs}(d).
There is an equal number of decagon rods and $\Delta$-chains in the VBC structure, as can be inferred from their mutual arrangement in the $[110]$ planar projection displayed in Fig.~\ref{fig:gs}(b).
The two building blocks are reminiscent of i) the perfect hexagon network and  ii) the pinwheels of the Marston-Zeng VBC~\cite{mz,sh} on the kagome lattice.

The 36 dimers per unit cell of the crystal can be assigned as follows: 18 form links between empty triangles, out of which 16 belong to the two decagon-rods while two connect
rods in only one of the two directions transverse to the rods, forming two dimensional layers within in the three dimensional lattice as shown in Figs.~\ref{fig:gs}(a,b).
Out of the remaining 18 dimers, 12 are located on the $\Delta$-chains, while the other 6 shield the $\Delta$-chains from the network of empty triangles located on the decagon rods.

A decagon-rod can be formed along a backbone [indicated by the line of blue and red bonds in Fig.~\ref{fig:gs}(c)] in either of the six equivalent $[110]$ directions inherited from the straight lines of bonds of the underlying pyrochlore lattice before depletion. It turns out that choosing a backbone and one of two possible dimer covering thereof fixes the rod and subsequently the entire VBC completely (up to the individual two-fold degeneracy of the $\Delta$-chains), which implies a total of 72 distinct VBC ground states.

\paragraph{Finite temperature ordering transitions ---} 
The VBC we have described is of course not the first one to be proposed for frustrated quantum magnets which exhibits a large unit cell with a complex ordering pattern---see proposals for kagome \cite{mz, nikolic} or pyrochlore \cite{berg,msg,hbb,tsunet}. How is one to verify or falsify these proposals? This is all the more pressing since one is unlikely to encounter the `perfect' Hamiltonian,  Eq.~\eqref{eqn:ham}; nor will it be easy to probe any given material below a potentially tiny ordering temperature. Can one nonetheless make some qualitative predictions?

With this motivation, let us discuss the behaviour of the magnet as $T$ is raised. Whereas it is in principle conceivable for there to be a single transition into the paramagnet, this is rather unlikely: there is no a 
priori reason why one of the different components of the complex ordered state should not melt before the others, leaving a partially ordered state behind. Our VBC involves for instance a $\mathbb{Z}_2$ degree of freedom encoding which sublattice of triangles hosts the triangles not occupied by a dimer (the triangles of the \hk lattice exhibiting a bipartite structure).
The expansion in $\lambda$ in fact suggests that this is the most robust feature, predicting a relatively high temperature Ising transition separated from other transition(s) which occur at lower temperature~\footnote{This scenario is supported by finite temperature Monte Carlo simulations of the classical dimer model~\eqref{eqn:eff_ham}, which clearly exhibits two finite temperature transitions, the first one conceivable at higher temperature being continuous and in the 3D Ising universality class, while the second one is of first order.}. The situation in the kagome \cite{mz} and pyrochlore \cite{berg,msg,hbb,tsunet} examples is entirely analogous.

\paragraph{Excitations and emergent low-dimensional structures.---}

The proposed VBC state features a huge number of low-lying singlet excitations. For instance, we expect singlet excitations involving local rearrangements of dimer motifs living at an energy scale of the order of
the couplings $\beta J \approx 1.9 \times 10^{-2} J$ and $\gamma J \approx 1.3 \times 10^{-2} J$ at $\lambda=1$. 
A subextensive number corresponding to the two distinct arrangements of dimers on each $\Delta$-chain remains degenerate until even higher order in perturbation theory.

\begin{figure}[t]
\centerline{\includegraphics[width=1.0\linewidth]{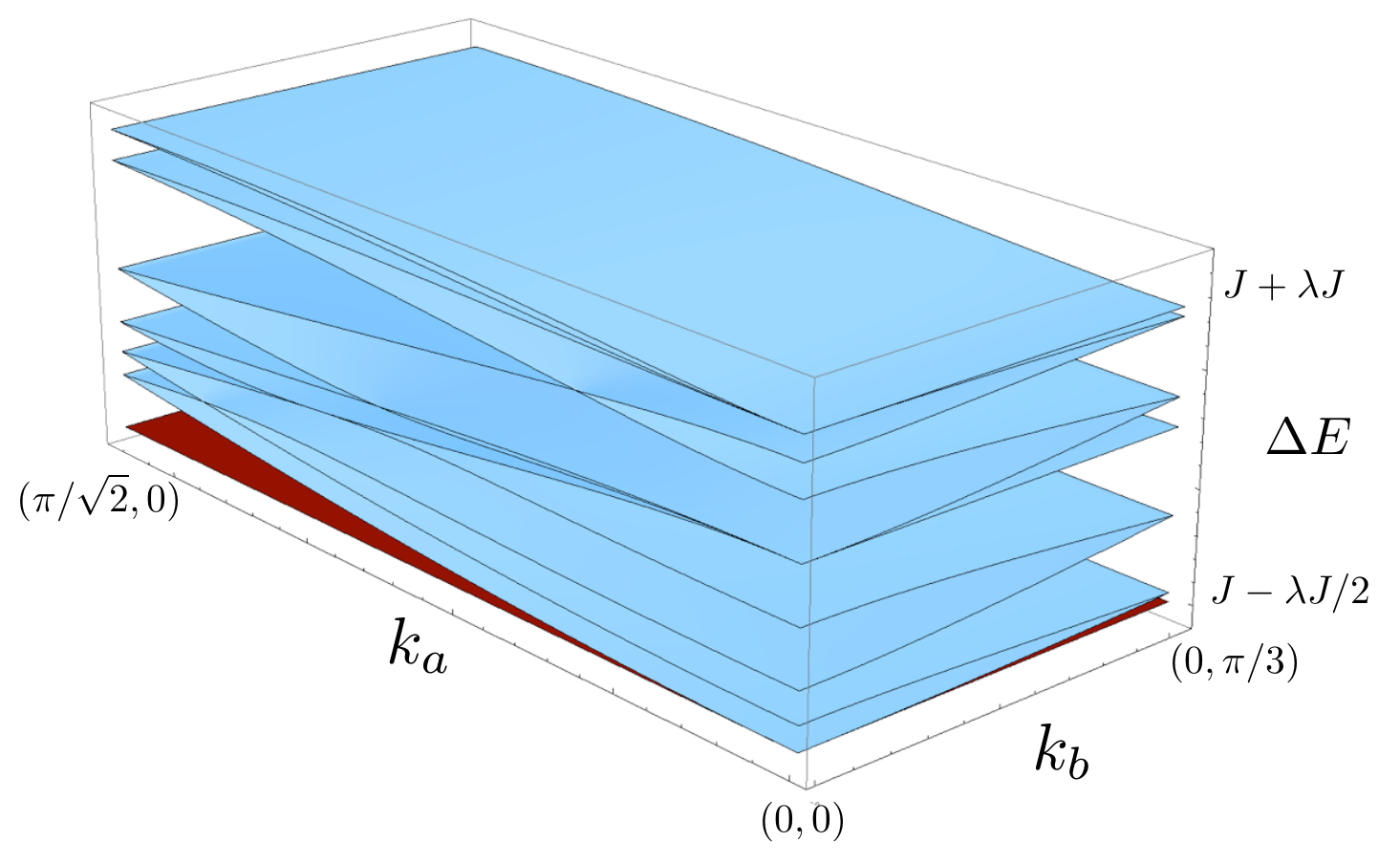}}
\caption{(Color online) Dispersion of the triplons to first order in $\lambda$, plotted as function of the momenta $k_a$ and $k_b$ in the decagon-rod plane spanned 
by the vectors $\mathbf{a}$ and $\mathbf{b}$ defined in Fig.~\ref{fig:gs}. The dispersion is flat in the perpendicular $k_c$ direction for all the modes. The inert triplets
at constant energy $J$ are not shown.
\label{fig:triplets}}
\end{figure}

The natural triplet excitation picture at small $\lambda$ is to excite a strong dimer from the singlet to a triplet state (called triplon) with an energy cost $J$. 
In first order perturbation theory this triplon will possibly delocalize by hopping processes between strong dimers. Analyzing the resulting tight-binding model of triplons
we find that triplons on the $\Delta$-chains and the shielding dimers (18 dimers in total) remain completely localized on their respective dimer due to 
the absence of hopping terms in first order.
The remaining 18 dimers in the VBC unit cell form 12 dispersive 
and 6 (lowest lying) flat triplon bands confined to a {\em two-dimensional} layer of decagon rods and their bridging bonds. The resulting dispersion is shown in Fig.~\ref{fig:triplets}.
Of the six triplons in the flat band (shown in brown in Fig.~\ref{fig:triplets}), four may be localized on the four perfect decagons in the VBC unit cell, while the remaining two on loops with 8 dimers involving the bridging dimers and extending beyond a single unit cell.

When going beyond the first order treatment we anticipate similar irregular behavior of the series as on the kagome lattice~\cite{sh2}. At any rate, the huge number of low lying excitations makes it seem very likely that coherent excitations will exist only in a very small window in energy-momentum space.

A complementary magnetic excitation picture has recently been developed in Ref.~\onlinecite{spinon}, based on the presence of gapped, deconfined spinons on the Husimi cactus 
(Bethe-lattice analogue of the kagome and \hk lattice) and the one-dimensional $\Delta$-chain. Since our VBC state explicitly features rather isolated $\Delta$-chains it is natural to 
assume that the completely localized triplons obtained in the first order treatment above actually decay into a spinon continuum when approaching the homogeneous $\lambda=1$ limit. 

Numerical simulations of the dynamical spin response on the kagome lattice~\cite{LL} suggest that a combination of both pictures might be appropriate.
At energies just above the spin gap the spectral density is quite high, reminiscent of the coherent response of triplon modes, while at
higher energies a rather featureless continuum is visible, reminiscent of the deconfined spinon picture.  
The confinement of triplon excitations to two-dimensional decagon-rod planes suggest that spectral function features are
almost momentum independent in the direction transverse to these planes.

{\it Discussion and relevance to experiment.---} 
In this Letter we have presented a candidate valence bond crystal ground state for the spin-$1/2$ Heisenberg model on the \hk lattice. The regularity and favorable energetics of our series expansion establishes the VBC as a serious contender to the earlier spin liquid proposals~\cite{lawler1,zhou,lawler2}. It would be interesting to investigate this further using the ``resonating'' approach of Ref. \onlinecite{kagomedimer}, which treats quantum dynamics in a different manner. 

To what extent a Heisenberg model is relevant for low energy physics of the recently studied Na$_4$Ir$_3$O$_8$ material is a matter of current research---its relevance has been argued for by several authors (see eg Refs. \onlinecite{exp,gang,micklitz}) and assumed by others, while there are also reasons to believe that the material is a weak Mott insulator undergoing a transition to a metal under pressure, thus making the assumption of a nearest-neighbor Heisenberg description thereof doubtful (see {\it eg} Refs.~\onlinecite{mott1,mott2,balents}). However, we note that the lack of magnetic ordering as well as the existence of a peak in $C_V/T$ in the experiment is a priori consistent with our picture. We also suggest that an anisotropic spin structure factor arising from the low-dimensional excitations discussed above serve as a further probe of our theory.

We acknowledge useful discussions with F.~Alet, Z.~Hao, D.~Huse, K.~Penc, I.~Rousochatzakis and H.~Takagi.

\end{document}